\documentclass[conference]{IEEEtran}
\IEEEoverridecommandlockouts
\usepackage{cite}
\usepackage{amsmath,amssymb,amsfonts}
\usepackage{algorithmic}
\usepackage{graphicx}
\usepackage{textcomp}
\usepackage{comment}

\usepackage{xcolor}
\usepackage[colorlinks]{hyperref}
\hypersetup{
     colorlinks   = true,
     linkcolor    = blue,
     urlcolor     = blue,
     citecolor    = blue
}
\usepackage{flushend}
\def\BibTeX{{\rm B\kern-.05em{\sc i\kern-.025em b}\kern-.08em
    T\kern-.1667em\lower.7ex\hbox{E}\kern-.125emX}}
    
\def\TODO#1{{\color{red} #1}}
\def\myparagraph#1{{\noindent\textbf{#1:}}}
\begin{document}

\title{Improving scalability and reliability of MPI-agnostic transparent checkpointing for production workloads at NERSC\\
}

\author{
\IEEEauthorblockN{Prashant Singh Chouhan}
\IEEEauthorblockA{\textit{Northeastern University}\\
Boston, USA \\
chouhan.p@northeastern.edu}
\and
\IEEEauthorblockN{Harsh Khetawat}
\IEEEauthorblockA{\textit{North Carolina State University} \\
Raleigh, USA \\
hkhetaw@ncsu.edu}
\and
\IEEEauthorblockN{Neil Resnik}
\IEEEauthorblockA{\textit{Northeastern University}\\
Boston, USA \\
resnik.n@northeastern.edu}
\and
\IEEEauthorblockN{Twinkle Jain}
\IEEEauthorblockA{\textit{Northeastern University}\\
Boston, USA \\
jain.t@northeastern.edu}
\and
\IEEEauthorblockN{Rohan Garg}
\IEEEauthorblockA{\textit{Nutanix, Inc.} \\
 Seattle, USA \\
rohan.garg@nutanix.com}
\and
\IEEEauthorblockN{Gene Cooperman}
\IEEEauthorblockA{
\textit{Northeastern University}\\
Boston, USA \\
gene@ccs.neu.edu}
\and
\IEEEauthorblockN{Rebecca Hartman--Baker}
\IEEEauthorblockA{
\textit{Lawrence Berkeley Nat. Lab.} \\
Berkeley, USA \\
rjhartmanbaker@lbl.gov}
\and
\IEEEauthorblockN{Zhengji Zhao}
\IEEEauthorblockA{
\textit{Lawrence Berkeley Nat. Lab.} \\
Berkeley, USA \\
zzhao@lbl.gov}
}

\maketitle

\begin{abstract}

Checkpoint/restart (C/R) provides fault-tolerant computing capability, enables long running applications, and provides scheduling flexibility for computing centers to support diverse workloads with different priority. It is therefore vital to get transparent C/R capability working at NERSC.
MANA~\cite{garg2019mana}, a transparent checkpointing tool, has been selected due to its MPI-agnostic and network-agnostic approach. However, originally written as a proof-of-concept code, MANA was not ready to use with NERSC’s diverse production workloads, which are dominated by MPI and hybrid MPI+OpenMP applications.
In this talk, we present ongoing work at NERSC to enable MANA for NERSC’s production workloads, including fixing bugs that were exposed by the top applications at NERSC, adding new features to address system changes, evaluating C/R overhead at scale, etc. 
The lessons learned from making MANA production-ready for HPC applications will be useful for C/R tool developers, supercomputing centers and HPC end users alike.
\end{abstract}


\section*{}

Transparent checkpointing for HPC has been discussed and developed at least since the 1990s~\cite{litzkow1997checkpoint,barak2013mosix}.  Yet, it is not in common use at the level of supercomputing.  This work describes an effort to make transparent checkpointing available for HPC production workloads:  first for users at NERSC~\cite{NERSC}, and then as a demonstration of best practices for other supercomputing sites to benefit from.

The work is based on the use of MANA for MPI~\cite{garg2019mana}, which in turn is based on the DMTCP package for transparent checkpointing~\cite{ansel2009dmtcp}. MANA employs a new split-process model, first described in 2019.  While that initial prototype demonstrates the practicality and advantages of the split-process approach, it has not yet been made production-ready for supercomputing.

In order to place MANA in context, the evolution of transparent checkpointing toward support for supercomputing is briefly summarized in the figure below.  (``Checkpointing'' or ``checkpoint-restart'' will often be abbreviated to ``C/R'' in the rest of this article.) 

\begin{figure}[ht]
\centering
\fbox{\begin{minipage}{0.9\columnwidth}
\centerline{\emph{A Short History of Checkpointing for MPI}}
Early: Application-specific checkpointing is widely used \\
1990s: Single-computer checkpointing \cite{litzkow1997checkpoint,barak2013mosix} \\
2005: BLCR: Berkeley Laboratory C/R~\cite{hargrove2006blcr} \\
2009: Interconnect Agnostic C/R for Open MPI~\cite{hursey2009interconnect} \\
2014: Transparent C/R for InfiniBand (below MPI)~\cite{CaoEtAl14} \\
2016: Petascale-level transparent checkpointing~\cite{cao2016system} \\
2019: MANA: MPI-Agnostic, Network-Agnostic~\cite{garg2019mana} \\
2020: Production-ready MANA (this work)
\end{minipage}
}
\end{figure}

The ``$M\times N$'' problem is the problem of supporting $M$~possible variants of MPI and $N$~possible variants of network. Supercomputing centers generally procure new systems every 3 to~5 years, which may use completely different MPI or network architectures. Therefore any transparent checkpointing solution must solve the ``$M\times N$'' problem to ensure portability from one generation of machine to the next. The MANA architecture was particularly promising in this regard.

MANA solves the ``$M\times N$'' problem through a \emph{split-process} approach.  The memory regions of the MPI application are tagged as \emph{upper-half} regions, and the MPI, network and other system libraries are tagged as \emph{lower-half}.  Only the upper half is checkpointed.  On restart, a trivial MPI application is created, thus instantiating the lower half.  Each MPI rank of this trivial application then restores the upper-half memory regions of that same rank. For details on DMTCP~\cite{ansel2009dmtcp} and MANA~\cite{garg2019mana}, see the relevant citations.

This work is an effort to improve the scalability and reliability of MANA. Although the prototype version was functional, it was nowhere close to supporting diverse production workloads running at all scales. This work has largely increased the scalability, reliability and usability of the software, and has also added new features that are required for the code to function after system upgrades. It can now reliably C/R a couple of top applications at NERSC. Further, the performance overhead of the software was evaluated over various file systems available on Cori~\cite{cori}, to prepare for MANA deployment at large scale.

\subsection*{NERSC Production Workloads}
NERSC is the primary HPC center for US DOE Office of Science, supporting more than 8,000 users and 900 projects from all scientific fields. Tens of thousands of different application binaries, which are from MPI or MPI+OpenMP codes dominantly, run at NERSC each year; jobs run at all scales --- from single node to full machine. 

Checkpoint/restart provides fault-tolerant computing capability, enables long running applications, and provides scheduling flexibility to support diverse workloads with different priority levels, e.g., making space for high-priority, real-time workloads by preempting low-priority jobs. It is therefore vital to get transparent C/R capability working at NERSC.

\begin{figure}[ht]
\centering
\includegraphics[width=0.8\columnwidth]{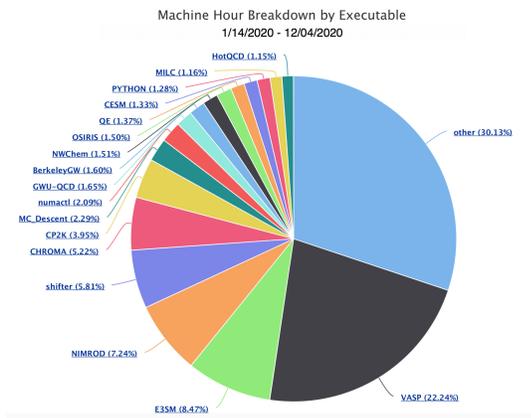}
  \caption{Application Usage at NERSC in 2020}
  \label{fig:nersc-workloads}
  \vspace*{-0.5\baselineskip}
\end{figure}

While it is a daunting task to transparently checkpoint and restart a huge number of applications run at NERSC at the system level, this task can be broken into small, incremental steps, prioritizing top applications. As shown in Figure~\ref{fig:nersc-workloads} the top 20 applications account for about 70\% of NERSC's Cori~\cite{cori} computing cycles. If we can get MANA to work reliably with these top applications, then potentially about 70\% of the system resources can be preempted to support the high priority, real-time workloads. 

To get MANA to work with NERSC's production workloads at all scales, we needed to address a few challenges. How production-ready and scalable is this approach, and how much overhead does it create? Under this high-level concept, there are minute details that need to be taken care of 
while implementing MANA in any checkpoint software.

\subsection*{Software Design Issues at Small Scale}
When implementing MANA for NERSC workloads, we observed issues of varying levels of severity at different scales. We began debugging at small scales, where we faced primarily bugs related to synchronization, memory-tagging, and other implementation issues for split processes. To resolve these issues, we instrumented the code to add rank-to-node and process-id mapping for debugging.

 Some of the bugs were related to TCP sockets, network delays, missing locks, descriptor conflicts, and lost messages. The DMTCP coordinator connects to each rank via a TCP connection. Network congestion on the production machine at times caused packet losses and disconnects. The TCP KeepAlive option was added to solve this problem. Network delays due to quiescence of the Cray GNI network reconfiguring itself brought additional bugs to the surface. And a few race conditions were seen when the data structures were left in an inconsistent state due to missing locks. The descriptor conflicts would occur upon restart: the upper half opens a file descriptor before checkpoint, and upon restart the lower half opens the same file descriptor number for its internal use. During restart, the lower half then restores the upper half application, creating a file descriptor conflict. We resolved this contention by tagging and reserving file descriptors for each half --- upper and lower. And to ensure that no in-transit MPI messages are lost due to checkpointing, we delayed the final checkpoint until the count of total bytes sent and received was equal. 
 
 Further, the original MANA assumed that addresses of certain system memory regions were fixed. When the operating system on Cori was upgraded, these assumptions were no longer true, resulting in some memory-region overlaps. To resolve this issue, we used the MMAP\_FIXED\_NOREPLACE option with mmap to dynamically determine free memory space each time the application executes.

\subsection*{Software Design Issues at Large Scale}
At large scale, we started to see new memory corruption, network failure, argument length limit, and disk space errors. We also began to see startup time performance issues with our dynamically linked MANA/DMTCP executables, as static linking is preferred at scale.

We found that the MPI library (in the lower half) can create new memory regions for message exchange at runtime. These lower-half regions may overlap with upper-half regions, eventually leading to memory corruption. MANA converts blocking MPI calls (e.g., MPI\_Send) to non-blocking MPI calls (e.g., MPI\_Isend); without sufficient care, this subtle difference in calls can change the semantics of an application. The Slurm \texttt{srun} command uses a network packet containing the list of arguments it was passed, to send commands to its worker processes. Due to the limit on packet sizes, \texttt{srun} was unable to pass all checkpoint file names to its workers, leading to a crash. We resolved this by changing the way we provide the file names. Applications with a large memory footprint may fail to checkpoint if there is insufficient storage space for the checkpoint image; a system warning is needed.

For best startup performance at scale, it is recommended to broadcast a statically linked executable to all nodes. DMTCP currently does not support static linking, but we plan to use the \texttt{--wrap=symbol} flag of the Linux linker to interpose on important functions needed by MANA.

\subsection*{Checkpoint Overhead Evaluations}

\begin{figure}[ht]
\centering
\includegraphics[width=0.9\columnwidth,height=2.5in]{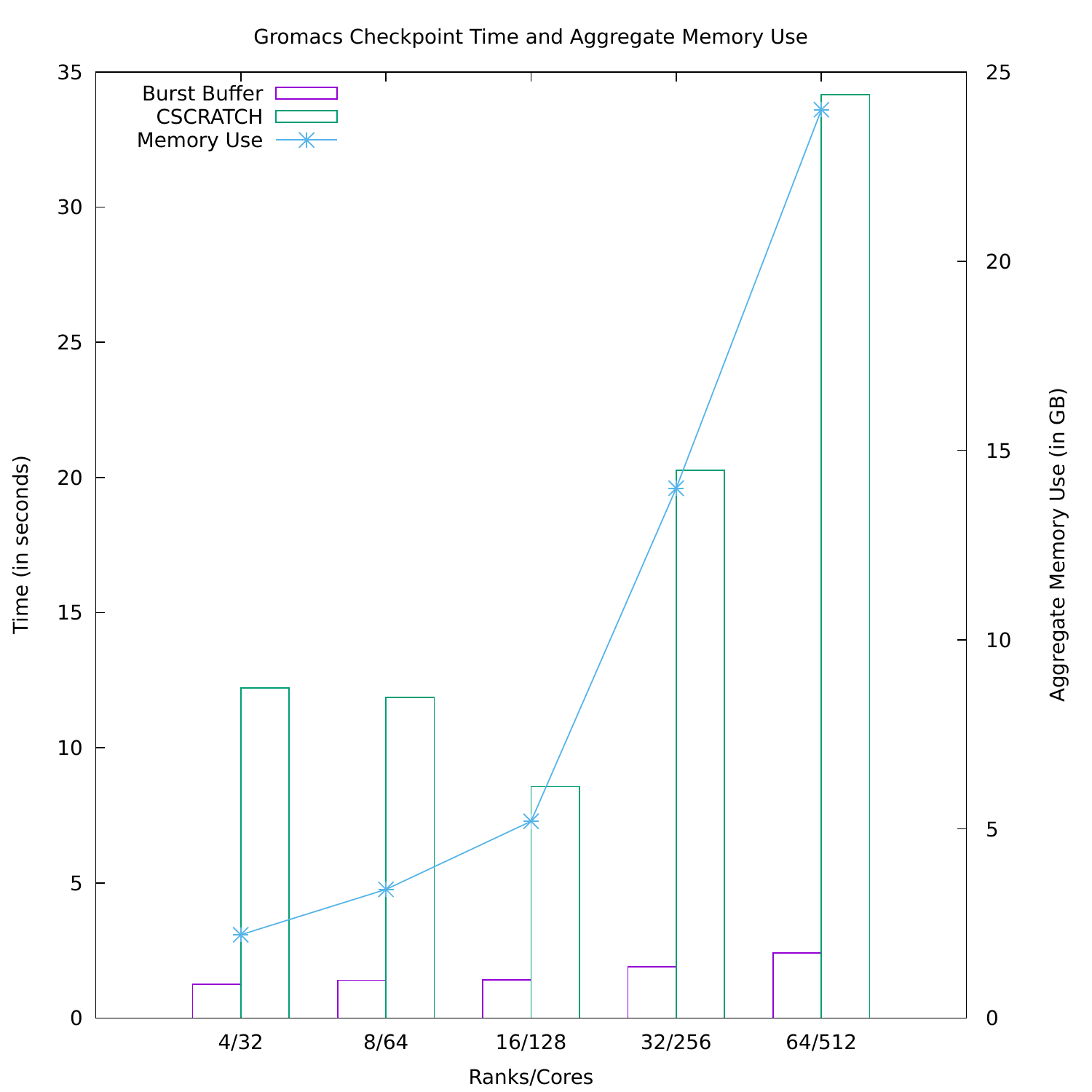}
  \caption{Checkpoint time of Gromacs with MANA on Burst Buffers and the Lustre file system CSCRATCH on Cori}
  \label{fig:gromacs-perf}
  \vspace*{-0.5\baselineskip}
\end{figure}

As part of ongoing work, we evaluated MANA's checkpoint overhead using multiple applications on different file systems. Figure~\ref{fig:gromacs-perf} shows the checkpointing time for Gromacs~\cite{GROMACS} for runs ranging from four ranks to 64 ranks with eight OpenMP threads per task using the ADH benchmark~\cite{GromacsBenchmark}. The aggregate memory use is shown in blue and average time for checkpoint is shown in purple and green, for Burst Buffers and the Lustre file system (CSCRATCH), respectively. Preliminary results show that performance on the Burst Buffers is superior to that on the CSCRATCH and also scales better. 

HPCG~\cite{HPCG} displays behavior similar to Gromacs, with checkpoint time for Burst Buffers at 30~seconds and CSCRATCH at over 600~seconds for 512 ranks with eight OpenMP threads per task. The aggregate memory used was 5.8~TB. The speedup for Burst Buffers over CSCRATCH on restart was more modest at about 2.5 times whereas the speedup for checkpointing was more than 20 times.

\subsection*{Lessons Learned}
Research codes are usually written quickly to demonstrate a proof of principle. While they show promising features, there is a gap to close before they can be used reliably in production. While debugging MANA and fixing bugs, we had to redesign a few parts of MANA. We learned that instrumenting research codes as follows would have greatly accelerated the conversion to production codes:
\begin{enumerate}
    \item Greater emphasis on runtime annotations. For example, an annotated table of all memory regions, along with dynamic runtime checks, would help catch bugs early in the development phase.
    \item Clearer semantic specifications for each code unit. This will enable better unit testing.
    \item Improved design for atomic data structures even for single-threaded code. Each data structure should include a field “CHANGES\_PENDING”, which would act as a lock.
    \item  Better attention to warnings and error messages from the beginning. This would help diagnose issues quickly.
\end{enumerate}

\subsection*{Current status of MANA adoption in production workloads}
Currently we have enabled MANA with NERSC's top application,  VASP~\cite{VASP}, which is a widely used materials science code and represents more than 20\% of computing cycles at NERSC (Figure~\ref{fig:nersc-workloads}). VASP is written  in  Fortran 90 and parallelized with MPI (version 5) or MPI+OpenMP (version 6). VASP jobs usually run with a smaller number of nodes for a long time. While some features implemented in VASP, such as atomic relaxations, have internal C/R support, some other features,  such as Random Phase  Approximation  (RPA), have no such support. The RPA  jobs can run  for much longer than 48 hours, the max walltime allowed on Cori. In the past  we  had  to  make  special  reservations  for  these  jobs, now they can run on Cori by checkpointing/restarting with MANA. MANA has been tested with various representative VASP workloads and is ready for production deployment.

We have also enabled MANA for Gromacs, a  widely used molecular dynamics simulation code, for which we had to fix multiple bugs exposed only by production workloads at scale. Gromacs is written in C++ and parallelized with MPI+OpenMP. While Gromacs has  internal  C/R support,  a  benefit  of  MANA  is  that  a  Gromacs computation can be checkpointed at any point in its execution and resumed  to  generate  exactly   the   same   results   as   an uninterrupted run.

Currently we are in the process of enabling MANA on the rest of the top applications at NERSC using the real use cases provided by NERSC users. 

\subsection*{Future Work}
The future work includes: getting MANA to work with more applications, reducing the checkpoint overhead for large-scale applications, deploying a preempt queue for real-time workloads, and enabling MANA for NERSC's next pre-exascale computer, Perlmutter~\cite{Perlmutter}, an NVIDIA GPU system.  

\section*{Acknowledgement}
The authors would like to thank Steve Leak and Chris Samuel at NERSC for valuable discussions and help. We would also like to thank the reviewers for their valuable comments and feedback. This work was supported by the Office of Advanced Scientific Computing Research in the Department of Energy Office  of  Science  under  contract  number  DE-AC02-05CH11231.
This work was partially supported by National Science Foundation Grant OAC-1740218 and a grant from Intel Corporation.

\bibliographystyle{ieeetr}
\bibliography{supercheck-mana}

\vspace{12pt}

\end{document}